# Enabling Effective FPGA Debug using Overlays: Opportunities and Challenges

Fatemeh Eslami[1], Eddie Hung[2], and Steven J.E. Wilton[1]

[1]Department of Electrical and Computer Engineering
University of British Columbia, Vancouver, Canada
{feslami,stevew}@ece.ubc.ca

[2]Department of Electrical and Electronic Engineering
Imperial College London, London, U.K.
e.hung@imperial.ac.uk

***Abstract*—FPGAs are going mainstream. Major companies that were not traditionally FPGA-focused are now seeking ways to exploit the benefits of reconfigurable technology and provide it to their customers. In order to do so, a debug ecosystem that provides for effective visibility into a working design and quick debug turn-around times is essential. Overlays have the opportunity to play a key role in this ecosystem. In this overview paper, we discuss how an overlay fabric that allows the user to rapidly add debug instrumentation to a design can be created and exploited. We discuss the requirements of such an overlay and some of the research challenges and opportunities that need to be addressed. To make our exposition concrete, we use two previously-published examples of overlays that have been developed to implement debug instrumentation.**

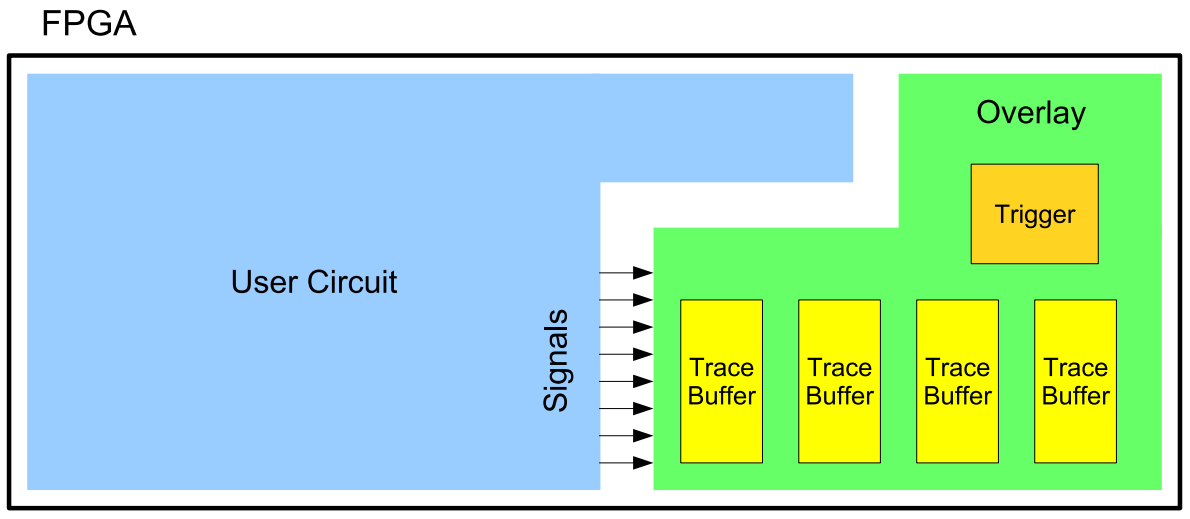

Figure 1: Overview of our approach

## I. Introduction

The past several decades have seen tremendous growth in the capacity and capability of Field-Programmable Gate Arrays (FPGAs). Today, FPGAs are poised to enter the mainstream as compute accelerators, as evidenced by Intel's recent acquisition of Altera and Microsoft's public efforts to bring FPGA technology into the cloud [1]. For FPGAs to be successful in this new role, an entire design ecosystem is required. Traditional hardware designers may be willing to accept long design and debug cycles, however, application designers using FPGA technology to accelerate software applications may not. These designers may expect software-like compile times, and similar support for debug and optimization. In recent years, the concept of an *overlay* has emerged as a promising technology to provide this capability, and may become key to ensuring that FPGA technology is successful as it moves to the mainstream.

Several researchers have described how overlays can accelerate the design and compile time of FPGA designs, either providing an embedded processor-style fabric which can be programmed using software [2], [3], or a flexible fabric and a compilation infrastructure that can quickly map circuits to the fabric [4], [5], [6], [7]. In some cases, the infrastructure is specifically optimized for accelerator-type circuits [8] or collections of small processing units [9].

Compiling a design, however, is only half the battle [10]. Designers also need an effective mechanism to debug and optimize their designs. Although many bugs can be found through simulation, many of the most elusive and troublesome bugs can only be found by running the design on an actual FPGA. When incorrect behaviour in a running chip is observed, finding the root cause of the behaviour is complicated by a lack of *observability* and *controllability*. Controllability and observability can be added by including commercial or academic debug instrumentation [11], [12], [13], [14], [15], [16], [17], [18]. This instrumentation often records the run-time behaviour of selected signals in the chip, allowing it to be played back later using debug tools. Some degree of controllability is also provided; in [18], the user can single-step through code and set breakpoints.

Most of these debug flows, however, require the design to be recompiled every time the instrumentation is changed. For very large designs, this can be prohibitive (often called a "go home event") which can severely limit debug productivity. Further, recompiling a design may often lead to slightly different timing behaviour which may cause a bug to disappear or change. Incremental compilation techniques may accelerate compilation and reduce timing variability. However, because these are general-purpose techniques — in that they are designed to accelerate changes being made to the original circuit, as opposed to simply adding new read-only instrumentation — this can still be slow, especially if significant changes to the debug instrumentation are made.

Overlays can provide a solution. An overlay can provide a flexible, adaptable, but generic fabric which can be compiled with the design once, and then used to implement debug instrumentation as shown in Figure 1. As the instrumentation is changed at debug time, the overlay can be reconfigured rapidly without changing the underlying user circuit. Unlike a typical overlay, this overlay can be optimized not only for debugging-type applications, but also for the underlying user circuit on a circuit-by-circuit basis.

In this paper, we will describe the use of overlays optimized to implement debug instrumentation. We will argue that our approach can lead to significant debug productivity improvements, by providing fast debug turns increasing the ability of a designer to understand what is happening in his or her circuit. We will describe our vision of the overall flow, as well as elaborate on many of the research challenges that must be addressed before this approach can become a reality. To make our arguments concrete, we will summarize some of our previously published early work. We believe that overlays are uniquely suited to the debug task, and we hope that, through this paper, we can motivate the community to perform further research in this area.

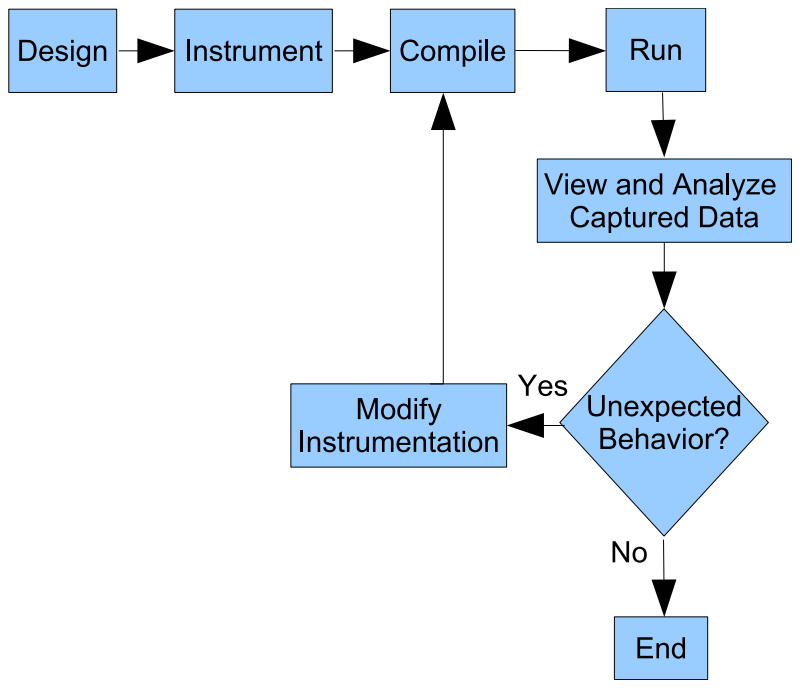

Figure 2: Typical debug flow

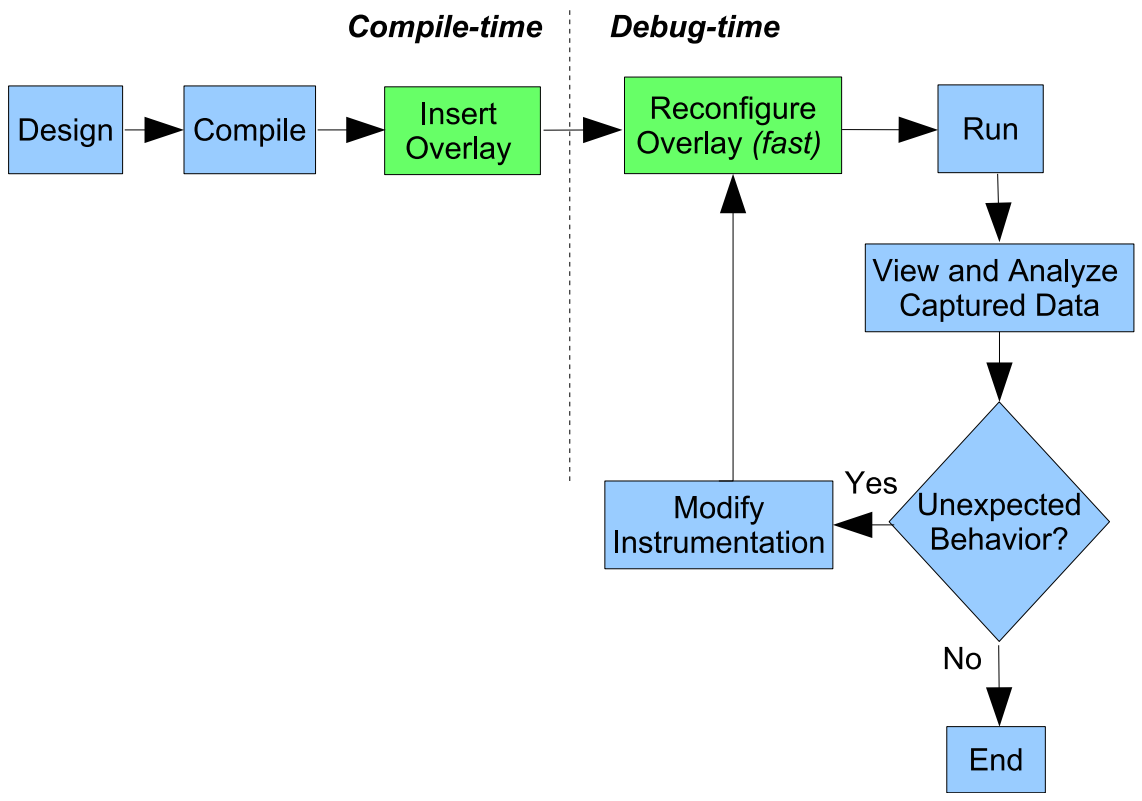

Figure 3: Debug flow of using an overlay for instrumentation

## II. Overlays for Debug

### A. Context

Although simulation can be used to uncover the root cause of many bugs, it is not enough. Many errors can only be observed when a design is running on the actual FPGA for several reasons: (1) many bugs only emerge after long run-times, and simulation is orders of magnitude slower than real silicon, (2) many "corner case" behaviours may need real (i.e. those that are impractical to describe with a test-bench) workloads before they can be observed, and (3) the most difficult bugs are often in the interfaces between a design and neighbouring chips; only by testing the system *in-situ* can such bugs be found.

Fundamentally, searching for the root cause of a bug requires understanding the circuit's behaviour. In simulation, the engineer can observe unlimited waveforms to help understand what is happening, however, in a running chip, there are limited I/O ports, meaning only selected signals can be brought outside the chip for observation. To increase observability, designers regularly use commercial tools such as Chipscope, SignalTap II, or Certus [11], [12], [13] to store the behaviour of important signals in on-chip memory. After execution, this data is read out, and used in a custom tool to replay the behaviour, hopefully leading to insight into the operation of the circuit. Academic solutions have also emerged [14], [15], [16], [17], [18]. During the hunt for the root cause of a bug, the designer often needs to select a different set of signals to record as he or she refines his or her view of how the circuit is operating. Similarly, the designer may need to adjust the trigger conditions so as to record behaviour in a different portion of the execution. Although these commercial solutions provide some level of configurability after insertion, for any significant changes, a recompile of the user circuit along with the instrumentation is often required. This can be very slow. Figure 2 shows a typical debug flow.

### B. Overlays: Our Vision

Overlays can be used to implement debug instrumentation as shown in Figure 1 using the flow in Figure 3. The overlay is a fabric that sits on the FPGA along-side the user circuit (which may be implemented on an overlay itself, or may be mapped directly to the native FPGA fabric). The debug overlay is compiled either at the same time or after the user circuit. At debug time, the overlay is configured to implement the desired debug function; for example, it may be configured to connect a set of signals of interest to trace buffers, or it may be configured to stop the execution at a predetermined breakpoint. After execution of the chip, data stored in the debug instrumentation can be extracted and used in conjunction with an off-line software tool to enable the user to understand the behaviour of the design. As the user refines his or her view of what is happening in the chip, he or she can reconfigure the overlay to set up a new debug scenario. It is important that the overlay be reconfigured without modifying (or recompiling) the user circuit, both to decrease the time between debug iterations as well as to ensure that small changes in the timing of the user circuit do not cause bugs to disappear.

The functions that can be implemented in the overlay can vary. Some of the more interesting ones are described below:

*Storage:* The heart of any debug instrumentation is a collection of memories (called *trace buffers*) which can be used to store the history of selected signals. FPGAs contain on-chip memories, so it is straightforward to implement these trace buffers on the device. These memories may be wide enough that each signal of interest can be written every cycle; alternatively, if the temporal nature of signal updates is known, a multiplexing circuit can be built to write different internal signals into the memory in each cycle [17].

*Interconnect:* In order to record the behaviour of selected internal signals in the design, connections must be made between the internal signals and the trace buffer memories. These connections would ideally be configurable, since different debug scenarios would require recording the behaviour of different signals. A straightforward implementation would be a configurable multiplexer or concentrator [19] that can connect a large number of potential signals to a smaller number of trace buffer inputs. A more interesting implementation of this interconnect will be described in Section IV.

*Triggering:* Since on-chip memory is limited, any practical implementation of a debug overlay will have an upper bound to the amount of data that can be stored in the trace buffers. Typically, trace buffers are implemented as circular buffers, meaning when a trace buffer fills up, older data gets evicted. Trigger circuitry is used to recognize a predetermined point during the execution of the design, and to stop recording data at that point. This allows the user to examine the behaviour of the design at a specific point (e.g. just before the bug surfaces) during the execution. In Section V we will describe one implementation of an overlay that supports implementing trigger circuitry; another approach is described in [14].

*Assertions and Monitors:* Rather than storing a complete history of a signal, it is sometimes preferable to create an assertion that triggers if a certain condition occurs [20], [21]. This logic can also be implemented using an overlay. Like a trigger circuit, an assertion would typically be a small combinational or sequential circuit. Although we do not describe an example of such an implementation in this paper, the trigger overlay described in Section V would likely be suitable for implementing assertions.

*Compression and Pre-Processing:* To make the most efficient use of trace buffers, it is often useful to compress data before it is stored. In general, trace data is extremely compressible. This compression may be general purpose as described in [22] or optimized for a specific application as in [17]. A general purpose compression engine may be too large to implement in an overlay, however, simple application-specific compression schemes may be very suitable. Similarly, it may be possible to process several signals in the overlay and store only the results; the manner in which data is processed can be configured by the user, and implemented in the overlay.

It should be pointed out that commercial solutions such as Chipscope or SignalTap II are, in essence, overlays themselves. In the examples presented later in this paper, we present overlays that are somewhat more flexible than the commercial offerings. In fact, there is a continuum in trade-off between overhead and flexibility in any overlay implementation; this will be discussed further in the next section.

## III. Research Opportunities and Challenges

### A. Construction

There are several ways the overlay architecture can be constructed. One option is to implement the overlay using normal FPGA logic resources. The overlay would be described in RTL, and compiled along with the user circuit as a single unit. The advantage of this approach is simplicity: the vendor tools can be used without modification, and the structure of the overlay can be easily modified and understood by the user. There are several disadvantages, however. First, compiling the fabric along with the user circuit means that the user circuit may be modified by the presence of the overlay network. This can be especially troublesome due to timing; if the critical path happens to be in the overlay network, the tool will not work as hard optimizing the timing of the user circuit. If the overlay is large, it is possible that the extra congestion slows *all* signals, including those in the user circuit. The fundamental problem is that the tool will treat the overlay architecture and user circuit as equally important and optimize both together.

An alternative is to place, route, and lock the user circuit, and *then* place the overlay network. This approach explicitly recognizes that the debug circuitry as being less "important" than the user circuit. In this way, the user circuit will be optimized on its own, as if the debug logic was not present, likely leading to more predictable user circuit implementations — since surely it would be better to observe fewer signals, than to obscure bugs entirely.

The second alternative can be taken further: rather than compiling an RTL description of the overlay fabric, and then mapping the debug circuit to this overlay, it can be far more efficient to map the debug circuit directly to an overlay constructed directly using available FPGA resources (LUTs, memories, and wiring segments). In the example of Section IV, the large multiplexers that form the connection between the user signals and trace buffers are constructed using the routing multiplexers within the switch blocks of the FPGA. Similarly, the overlay network in Section V is constructed using available LUTs. This technique avoids a downside of many overlay approaches: the high area overhead of implementing an "FPGA on an FPGA" is avoided. However, implementing such an overlay needs to employ custom incremental compilation tools. Developing enhanced compilation methods as well as finding ways to use the available FPGA primitives to implement debug logic is an interesting research direction that will enable much more efficient debug overlays.

### B. Architecture

Another critical factor that determines the effectiveness of our approach is the architecture of the debug overlay. Commonly, overlays are composed of coarse-grained units to minimize area overhead. If the incremental approach described in the previous section is used, however, a wider range of architectural possibilities can be considered. Indeed, the architecture in Section V resembles a fine-grained FPGA (but with reduced flexibility in the routing).

A unique opportunity exists to optimize the overlay network for a given user circuit. Since the user circuit is known when the overlay is constructed, it is possible to optimize the overlay architecture on a user-circuit by user-circuit basis. As an example, the structure of the circuit may give guidance into the type of assertions or monitoring circuitry that may be useful, and this may affect the granularity of the overlay logic resources, or the number and type of connections between the user circuit and the overlay. Investigating how an overlay network can be optimized on a circuit-by-circuit basis is an interesting research question that, if answered, could lead to significant capacity improvements.

An interesting research question resolves around the trade-off between flexibility and efficiency. Even using incremental techniques, coarse-grained overlays are likely to be more efficiently implemented on the FPGA fabric, as long as the debug circuit maps "nicely" to the coarse-grained architecture. One possibility may be to employ a combination of coarse and fined grained logic. Taken further, a very heterogeneous architecture would be an interesting approach.

The examples in Sections IV and V both assume the user circuit is not implemented on an overlay (but rather compiled to the native FPGA resources). If the user circuit is implemented on an overlay, another interesting architectural possibility exists: co-optimizing the overlay architecture for both the user circuit and the overlay network. This would eliminate the need for a separate debug overlay network, and would provide the ability to rapidly change the allocation of space between the user circuit and the debug instrumentation. Finding a way to create a single overlay architecture that works efficiently for the user circuit and the debug logic is an interesting research direction.

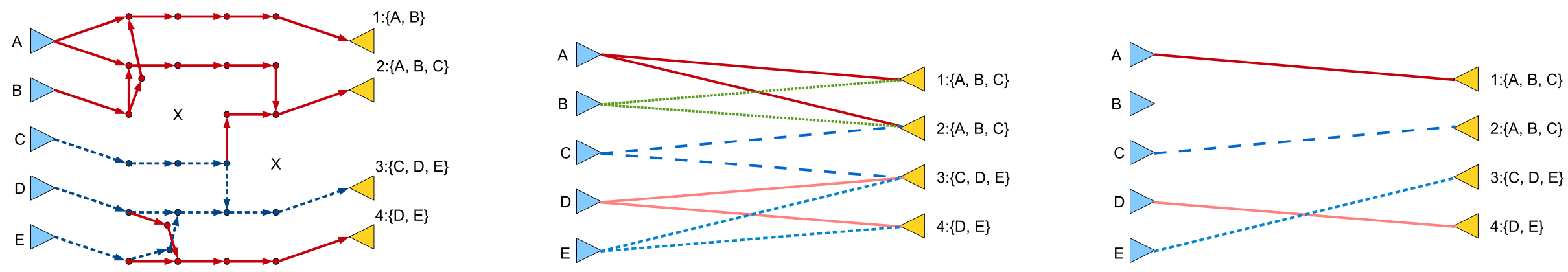

(a) Simple trace overlay network (b) Bipartite graph (c) Max. matching of graph

Figure 4: An example of using trace overlay for signal tracing, ▷: user signals, ○: routing multiplexers, ◁: trace buffer inputs.

### C. Computer-Aided Design Challenges

Designing a suitable overlay architecture is not enough: effective CAD algorithms are also critical.

There are two aspects to the compilation tools that need to be considered. The first stage, which we refer to as *overlay construction* is the mapping of the overlay fabric to the FPGA; this is performed once when the user circuit is compiled. The second stage, which we refer to as *debug circuit mapping* is the mapping of the debug circuit to the overlay fabric; this is performed every time the user wishes to change the debug logic. In both cases, the Quality of Results (QoR) is a critical factor in determining the viability of debug overlay networks. The run-time for debug circuit mapping is also important; the one-off run-time of the overlay construction less so (but it still must be controlled as to not increase the compile time of the original user circuit too much). In the example described in Section V, modified simulated annealing algorithms are used for both steps, however, other algorithms may also be appropriate.

A unique opportunity during the construction of the interconnect overlay is that we do not need to acheive 100% routability, as would exist in a fully-populated crossbar. If, during routing, some nets are difficult or impossible to route, these nets can be left out of the overlay fabric. It is then up to the debug circuit mapping algorithm to find a mapping that does not use any of the missing connections. This best-effort approach may potentially allow us to implement larger overlay fabrics that would otherwise be possible. The tradeoff between how aggressively we can prune "difficult connections" vs. how much this affects the routability of the overlay network is an interesting research question.

Finally, the tradeoff between architecture and CAD run-time is an important question. Different architectures (with different area overheads) may lead to simpler or more complex CAD problems. Understanding this tradeoff in the context of debug overlays would be an worthwhile research goal.

## IV. Example 1: Overlay Network for Signal Tracing

To make our ideas concrete, in this section, we describe an overlay network that allows designers to observe almost any subset of signals at debug time. For more details, see [23].

### A. Trace Overlay Architecture

The trace overlay architecture is used to implement connections between signals in the user circuit to trace buffer input pins in such a way that allows fast debug time reconfiguration. The aim is to connect all user signals (i.e. the output of all LUTs, FFs, RAMs, etc.) to at least one trace buffer input. To increase the flexibility of the overlay network, each signal can connect to multiple trace buffer inputs. A key opportunity exploited by this trace overlay is that in order for a user signal to be observed, the overlay needs only to be configured to route the signal to one of the many trace buffer inputs available, since they are all equivalent for this purpose.

### B. Trace Overlay Compilation

At compile time, after user circuit compilation, the overlay network is created out of spare routing resources left unused by the user circuit. We use a routing algorithm based on a modified version of PathFinder [24] to attempt to connect all user signals to a user-defined number of trace buffer inputs. Instead of using each routing multiplexer just once, we allow routing multiplexers to be overused, noting that their select bits can be determined at debug time when the overlay network is used for signal tracing. The result is a blocking network (more precisely, as shown in Figure 4a, a forest of trees whereby each tree is rooted at a trace buffer input, and its leaves are a subset of all user signals) in which a reduced amount of connectivity exists — we describe how this network is configured in the following section. Our VPR experiments showed that building an overlay network connecting 99.8% of all signals on top of the circuits targeting minimum-sized FPGAs with 30% more routing tracks than the minimum value increases the compile time by an average of 34%. However, this compilation time is only needed once to build the overlay architecture.

### C. Trace Overlay Configuration

At debug time, the designer chooses a subset of all user signals (up to the number of trace buffer inputs) to connect to the trace pins for observation. Due to limited flexibility of the overlay network, it may not be possible to connect all the requested signals simultaneously, since each routing multiplexer can only forward one signal from its input set. To make the decision of which signals to connect to the trace buffer pins, we create a bipartite graph which describes the possible connections between all nodes in the user circuit and all trace buffer pins as shown in Figure 4b (essentially, by folding the routing multiplexers from Figure 4a into the trace buffer inputs whilst preserving connectivity). We then utilize a maximum matching algorithm to find the solution in which the provably maximum number of requested signals can be connected to trace buffer inputs as shown in Figure 4c. Once this assignment is done, each routing multiplexer of the overlay is set to forward the assigned signal to the assigned trace buffer input. We showed any set of signals can be connected to 80-90% of the maximum trace buffer inputs in seconds. On average the circuit critical path delay increases by 9.0%, however this can be reduced significantly by pipelining the interconnect (increased latency between the user signals and trace buffer can be tolerated through post-processing before trace data is presented to the designer).

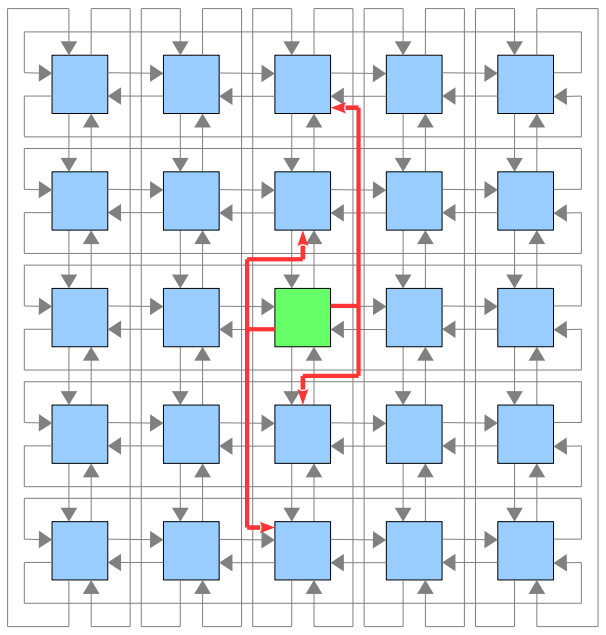

Figure 5: Simplified overlay architecture

## V. Example 2: Overlay Architecture for Trigger Insertion

Trigger insertion is more challenging than signal tracing since it involves implementing arbitrary logic functions. In this section, we describe a virtual overlay architecture that allows designers to insert trigger circuitry at debug time. For more details, see [25].

### A. Trigger Overlay Architecture

The architecture consists of a square grid of cells. Each cell contains between four and $N$ logic elements (LEs) where $N$ is the size of each cluster in the target FPGA architecture. To provide a reasonable tradeoff between area and flexibility, the cells are connected using a 2D torus, as shown in Figure 5. Each cell has twelve input pins and four output pins connecting it to the neighbouring cells. Each output signal is connected to three sinks: one *primary* sink and two *secondary* sinks. The primary sink connects the output pin to the one of the cell's nearest neighbour (e.g. to the north side). Although the primary sinks would be sufficient to provide complete connectivity between all cells in the overlay, additional routing flexibility is provided by the two secondary sinks. The first secondary sink of each output connects to a different nearest neighbour than the primary sink (e.g. the east side). The second secondary sink connects to a cell that is 1-hop away (not a nearest neighbour cell).

Some cells will contain more than four logic elements. Those logic elements do not drive any other cell; instead, they only drive other logic elements within the same cell through the local feedback crossbar. Cells that are not neighbouring can be connected by routing through one or more intermediate cells, and configuring these intermediate cells as pass-through buffers.

### B. Trigger Overlay Compilation

Here, we describe our CAD techniques for mapping the overlay architecture on top of the user circuit. We do not allow the rip-up and re-route of any user circuit nets, nor any relocation of any user circuit blocks.

*Cell selection:* Given the mapping of the user circuit, the algorithm first identifies all logic clusters inside the FPGA which contain *at least* four unused logic elements as well as at least 12 unused input pins. We then choose some subset of these logic clusters and create the largest possible square overlay out of the selected logic clusters. The result is a logical grid describing the overlay architecture.

*Adaptive placement:* At this stage, we perform a simulated annealing placement to position the cells in the overlay onto available sites in the fabric to optimize the total wirelength of the overlay wires. We only consider swaps which result in legal placements; a cell that consists of $b$ logic elements can only be moved to a site that contains at least $b$ empty LEs. This ensures that the final placement result is legal.

*Best-effort prioritized routing:* After overlay placement, we attempt to create all connections between the cells. Routability-driven PathFinder [24] is used to iteratively resolve routing congestion followed by a *best-effort* routing legalization heuristic. This heuristic iteratively discards illegal connections (from the overlay) until a legal routing solution is achieved.

We first use Pathfinder to route the *primary sinks* of each connection. We then use the same algorithm to route the *secondary sinks* of each connection. This approach ensures that the primary sink connections have higher priority for using routing resources than the secondary connections.

Our VPR experiments showed that building our overly architecture on top of the circuits targeting minimum-sized FPGAs with 30% more routing tracks than the minimum channel width increases the compile time by an average of 22%. However, this compile time is only required once to build the overlay architecture.

### C. Trigger Overlay Configuration

At debug time, the overlay architecture can be configured to implement a trigger function. The overlay architecture has limited flexibility in the routing network due to its simple topology and best-effort routing. Therefore, it is important that the placement algorithm is *routing-aware*.

*Routing-aware placement heuristic:* Simulated annealing is used to perform the placement of the trigger onto the overlay. Initially, each logic element of the trigger netlist is assigned to a random unoccupied logic element inside the overlay. LEs are swapped (as opposed to cluster-based placement as in [26]).

Figure 7 shows how a trigger netlist of Figure 6 can be mapped to overlay cells. LE1 and LE2 are placed in the same cell, sharing an input pin as well as connecting to LE4 and LE5 directly. Placing LE3 in the same overlay cell as LE1 and LE2 will result in routing failure; in case (1) the output pin of the LE of the overlay cell is only connected to the local routing crossbar and can not be connected to LE4. In case (2), although the output pin of the LE of the overlay cell can reach LE4, all input pins of C1 are occupied by LE1, and LE2, resulting in an unavailable input pin for LE3. Placing LE3 in C3, can indirectly connect to LE4 via C4. LE4 and LE5 connect to LE6 by the local crossbar, resulting a routable

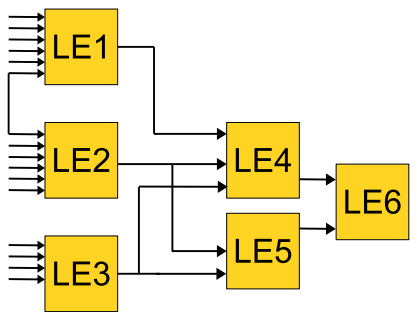

Figure 6: Simple trigger netlist

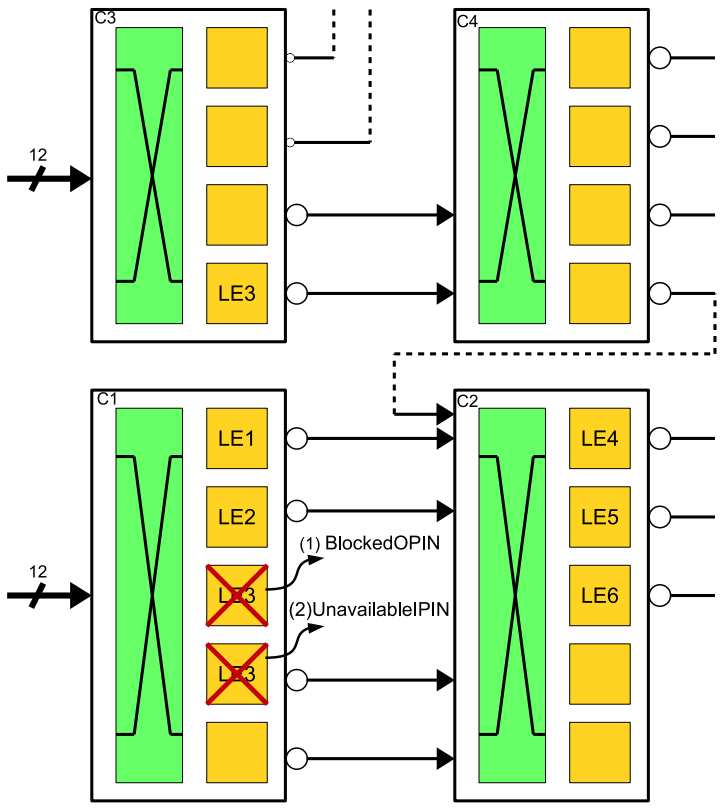

Figure 7: The trigger netlist of Figure 6 mapped to logic elements of overlay cells

trigger mapping.

To find a routable solution, swaps that result in indirect sinks are slightly penalized to encourage LEs to move to the same cluster or neighbouring clusters to make use of the intra-cluster connections or pre-routed connections of the overlay, respectively. Blocked output pins and unavailable input pins are heavily penalized since they will result in an immediate routing failure.

We showed trigger mapping is at least an order of magnitude faster rather than recompile insertion with negligible impact on delay.

## VI. Summary

One of the primary motivations for using overlays on FPGAs is that they can lead to fast compile times. Nowhere is this more important than in debug. Debugging often requires running a chip multiple times, each time with different instrumentation circuitry. Building this instrumentation on top of an overlay fabric provides the ability to quickly complete debug turns, significantly improving design productivity. Although some work has been done in this area, we feel there are many research challenges and opportunites that need to be addressed to fully exploit this technology. In this paper, we have discussed several of these challenges, along with some initial work to address some of them, however, there is still a lot more to do. We hope this paper has set the groundwork for what may become an extremely important research area.